\documentclass[aps,12pt,final,notitlepage,oneside,onecolumn,nobibnotes,nofootinbib,superscriptaddress,noshowpacs,centertags]
{revtex4-2}

\usepackage{multirow}

\usepackage{amsmath,amssymb}
\RequirePackage{mathtext}
\RequirePackage[utf8]{inputenc}
\usepackage{hyperref}
\usepackage{caption2}
\usepackage{graphicx}

\begin{document}

\title{The results of the development of\\the SPHERE-3 detector for studying\\the PCR mass composition in the 1--1000 PeV energy range.\\The status of 2025.}

\author{\firstname{D.\,V.}~\surname{Chernov}}
\email{chr@dec1.sinp.msu.ru}
\affiliation{Skobeltsyn Institute for Nuclear Physics, Lomonosov Moscow State University}
\author{\firstname{E.\,A.}~\surname{Bonvech}}
\email{bonvech@yandex.ru}
\affiliation{Skobeltsyn Institute for Nuclear Physics, Lomonosov Moscow State University}
\author{\firstname{O.\,V.}~\surname{Cherkesova}}
\affiliation{Skobeltsyn Institute for Nuclear Physics, Lomonosov Moscow State University}
\affiliation{Department of Space Research, Lomonosov Moscow State University}
\author{\firstname{E.\,L.}~\surname{Entina}}
\affiliation{Skobeltsyn Institute for Nuclear Physics, Lomonosov Moscow State University}
\author{\firstname{V.\,I.}~\surname{Galkin}}
\affiliation{Skobeltsyn Institute for Nuclear Physics, Lomonosov Moscow State University}
\affiliation{Faculty of Physics, Lomonosov Moscow State University}
\author{\firstname{V.\,A.}~\surname{Ivanov}}
\affiliation{Skobeltsyn Institute for Nuclear Physics, Lomonosov Moscow State University}
\affiliation{Faculty of Physics, Lomonosov Moscow State University}
\author{\firstname{T.\,A.}~\surname{Kolodkin}}
\affiliation{Skobeltsyn Institute for Nuclear Physics, Lomonosov Moscow State University}
\affiliation{Faculty of Physics, Lomonosov Moscow State University}
\author{\firstname{N.\,O.}~\surname{Ovcharenko}}
\affiliation{Skobeltsyn Institute for Nuclear Physics, Lomonosov Moscow State University}
\affiliation{Faculty of Physics, Lomonosov Moscow State University}
\author{\firstname{D.\,A.}~\surname{Podgrudkov}}
\affiliation{Skobeltsyn Institute for Nuclear Physics, Lomonosov Moscow State University}
\affiliation{Faculty of Physics, Lomonosov Moscow State University}
\author{\firstname{T.\,M.}~\surname{Roganova}}
\affiliation{Skobeltsyn Institute for Nuclear Physics, Lomonosov Moscow State University}
\author{\firstname{M.\,D.}~\surname{Ziva}}
\affiliation{Skobeltsyn Institute for Nuclear Physics, Lomonosov Moscow State University}
\affiliation{Faculty of Computational Mathematics and Cybernetics, Lomonosov Moscow State University}


\begin{abstract}
The SPHERE-3 setup is destined to register the EAS Cherenkov light, both the direct and the reflected from the snow surface. A set of methods and new approaches in the measurement technique enable substantial progress in the study of the primary cosmic ray composition in the energy range 1--1000 PeV. The present work reveals the current status of the project development and detector performance modelling.
\end{abstract}

\maketitle

\section{Introduction}

Initially, the SPHERE project aimed to implement the method of registering reflected Cherenkov light proposed by A.E. Chudakov~\cite{Chudakov1972}. This quasi-calorimetric technique allows for the most accurate estimation of the energy of the initial particle. In addition, this method is the most independent of models compared to other techniques for estimating energy. These advantages provide the prerequisites for more accurate determination of the characteristics of individual extensive air showers (EAS), as energy is one of two factors affecting the slope of the Cherenkov photons lateral distribution function of the EAS event. The other significant factor is the type of initial particle. This circumstance makes it possible to determine the chemical composition of primary cosmic rays (PCR) with greater accuracy. Therefore, a better understanding of energy allows for reducing the uncertainty in the reconstruction of the mass composition of the initial particles.

\begin{figure}[t!]
\setcaptionmargin{5mm}
\onelinecaptionstrue  
\includegraphics[width=0.5\linewidth]{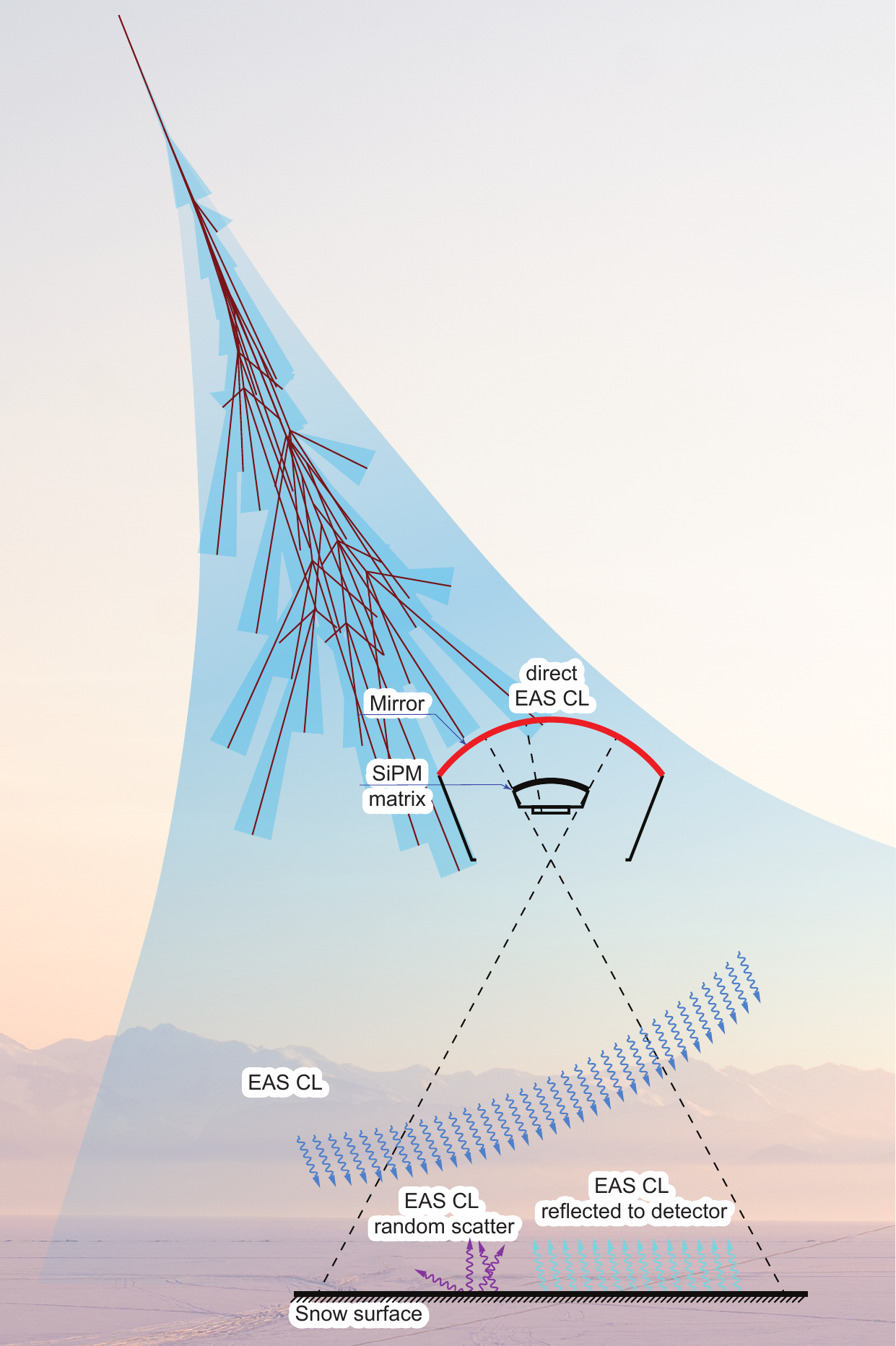}
\captionstyle{normal} \caption{Schematic representation of the SPHERE-3 project registration technique.}
\label{fig:scheme}
\end{figure}

In the Fig.~\ref{fig:scheme}, a schematic diagram of the SPHERE-3 experiment is provided. The method for recording reflected CL form EAS has been tested experimentally several times~\cite{Antonov1975,Navarra1981}, reflected CL was observed either from the mountainside or by lifting the installation above the snow-covered surface using a balloon. Due to advancements in modern technology, it is proposed to use unmanned aerial vehicle (UAV) as a carrier for this project.

The most significant and effective measurements were conducted on Lake Baikal between 2011 and 2013 using the SPHERE-2 detector. The energy spectrum was obtained, and the proportion of the light component was isolated~\cite{Chernov2017}. An original technique for isolating the light component and other nuclei groups was used~\cite{Antonov2015,Vaiman2021}. 

During the processing of the SPHERE-2 results, it was determined that searching for differences in heavy nuclei requires more complex multi-parameter criteria or additional information about the event. Previously, cases of registration of `parasitic' flashes from direct CL through gaps in the mirror segments have been found. This led to the idea of utilizing direct CL registration. Preliminary evaluations of the efficacy of direct CL recording in conjunction with reflected CL have yielded positive outcomes. The study revealed that analyzing the angular distribution of direct CL allows for events to be categorized into several mass groups. Consequently, conducting a comprehensive analysis of data regarding CL based on both reflected and direct has proven to be a highly promising endeavour.

\section{SPHERE-3 detector}

\begin{figure}[t!]
\setcaptionmargin{5mm}
\onelinecaptionsfalse 
\includegraphics[width=0.49\linewidth]{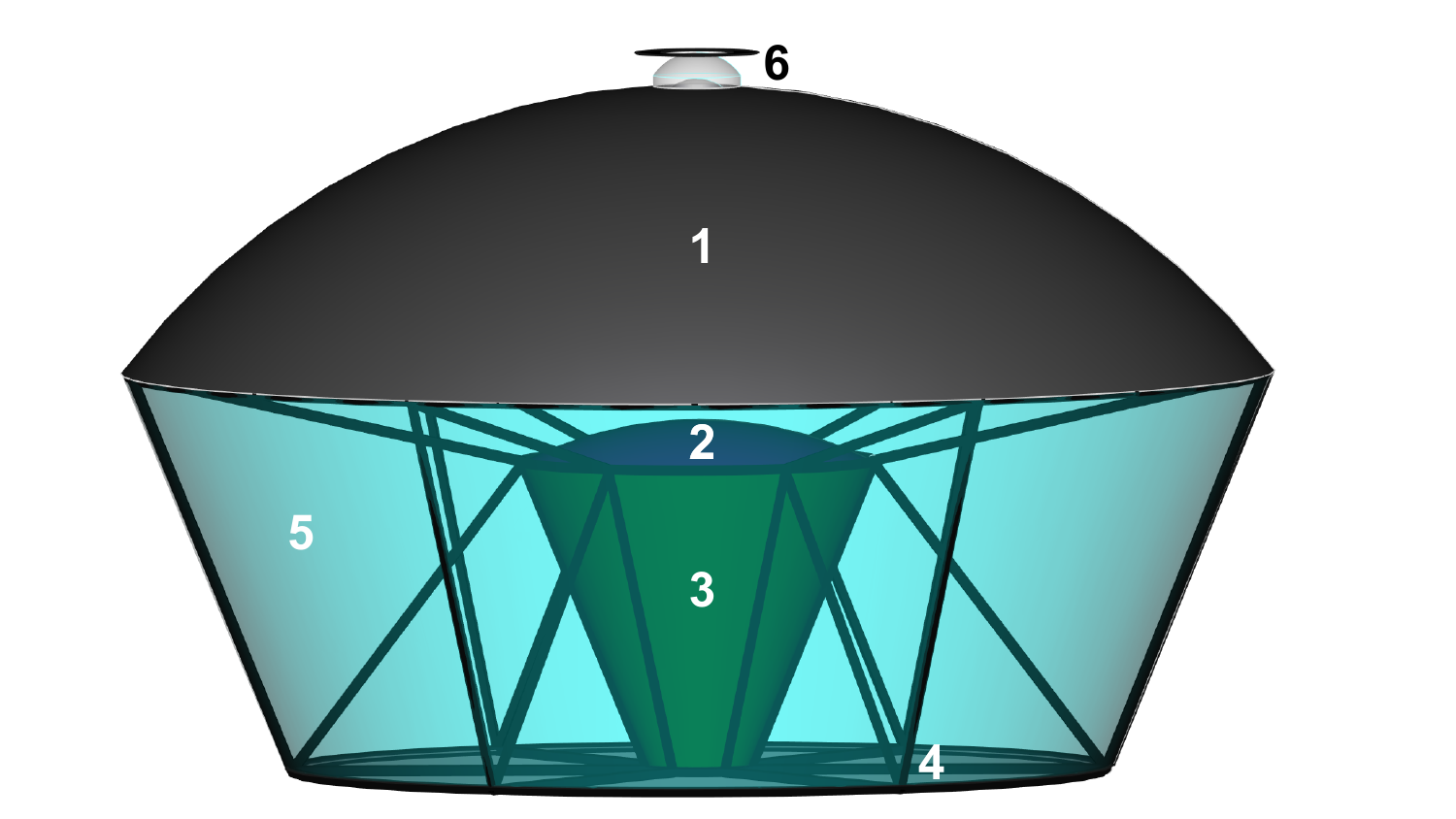}\hfill
\includegraphics[width=0.49\linewidth]{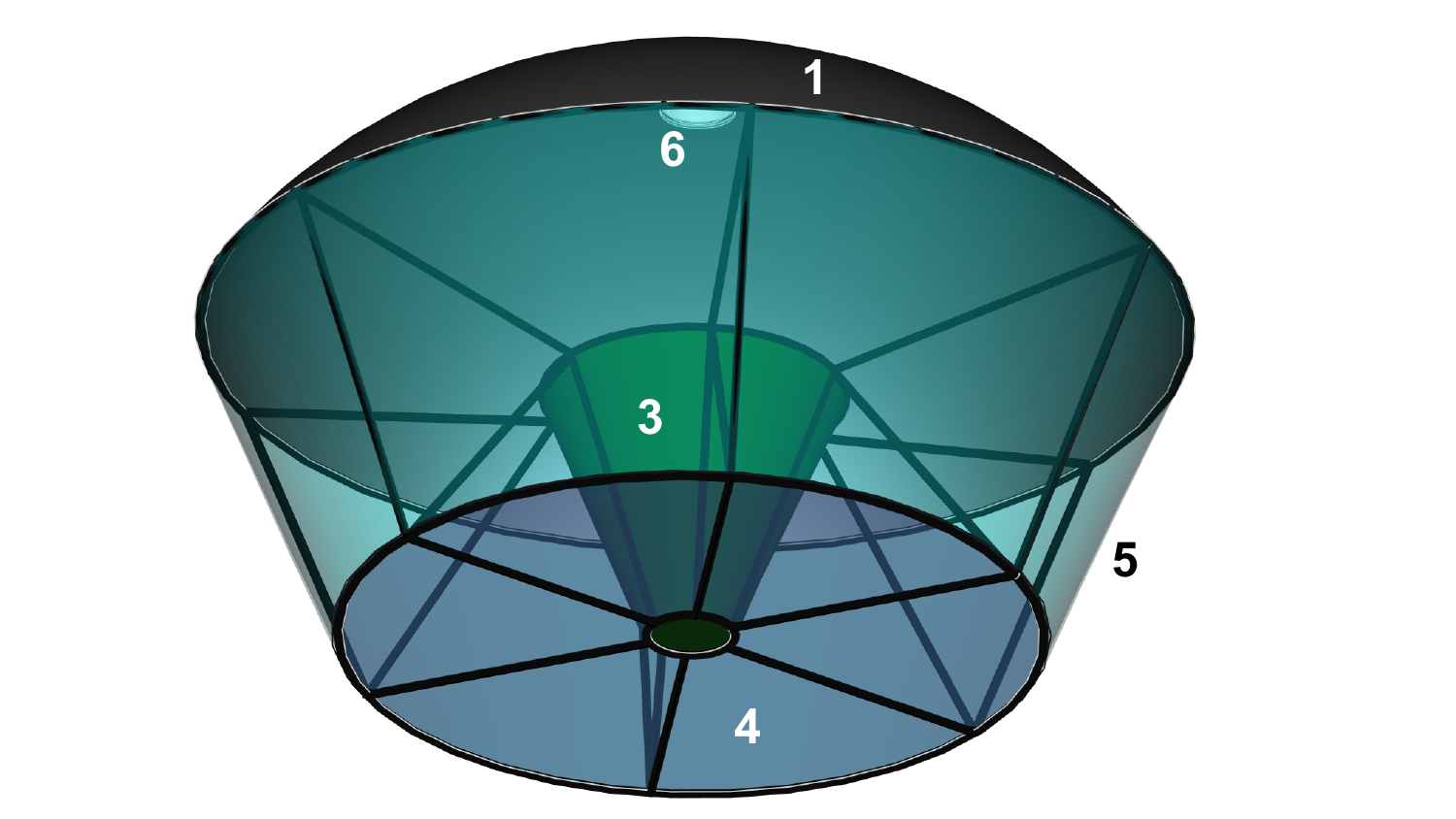}
\captionstyle{normal} \caption{Construction of the SPHERE-3 detector. 1 --- mirror, 2 --- SiPM mosaic, 3 --- electronic block, 4 --- corrector lens, 5 --- hood, 6 --- direct Cherenkov light lens.}
\label{fig:genview}
\end{figure}

In the Fig.~\ref{fig:genview}, a view of the main structural components of the SPHERE-3 system is provided. The Schmidt optical system is utilized in the telescope for reflected CL registration. An aspheric mirror composed of 7 or 19 elements (depending on manufacturer capabilities), with a total diameter of 2,200~mm, is utilized to collect light. The asphericity characteristics of this mirror are described by an 8th-degree polynomial, with the maximum deviation from a spherical surface being approximately 60~mm at the edge of the mirror.

To enhance the optical resolution of the detector a corrective lens with a diameter of 850~mm, is mounted on the entrance window. This lens also has an aspheric shape, described by a 16th-degree polynomial. The maximum thickness of this corrector is approximately 35 mm, and the minimum thickness is 10 mm.

\begin{figure}[t!]
\setcaptionmargin{5mm}
\onelinecaptionsfalse 
\includegraphics[width=0.49\linewidth]{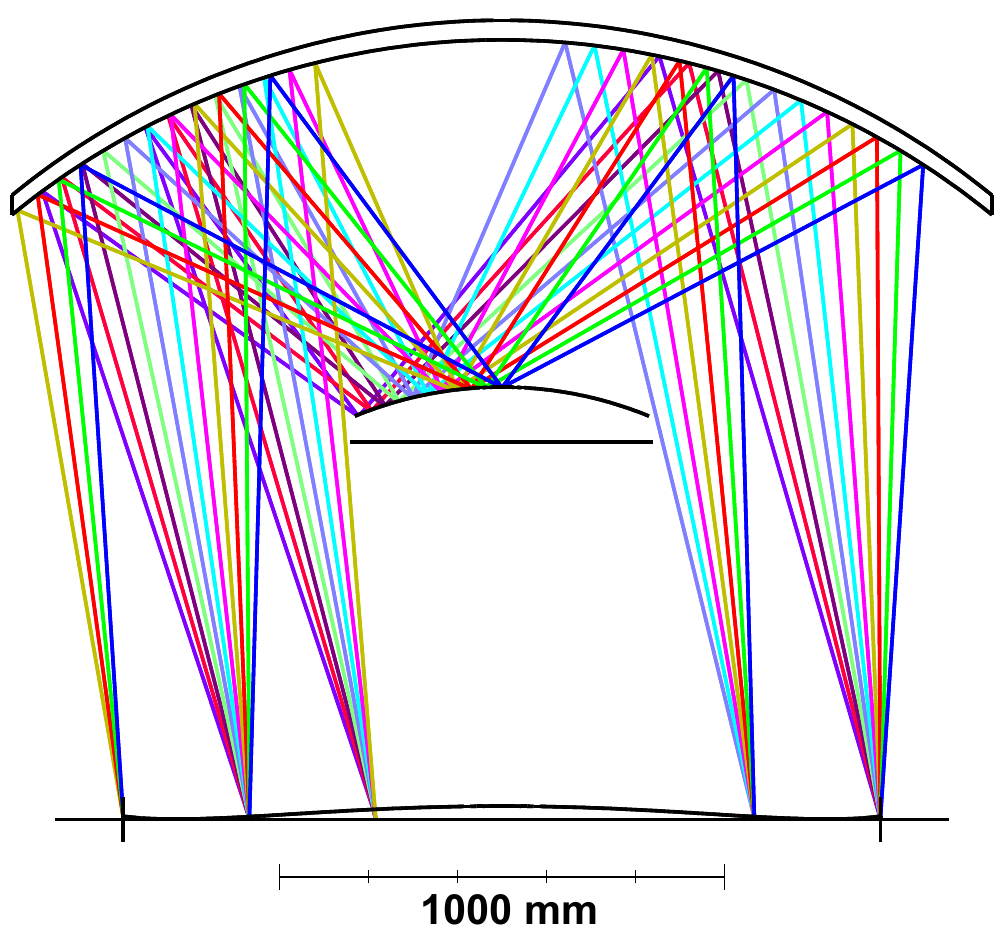}\hfill
\includegraphics[trim={0 -9cm 0 0},width=0.40\linewidth]{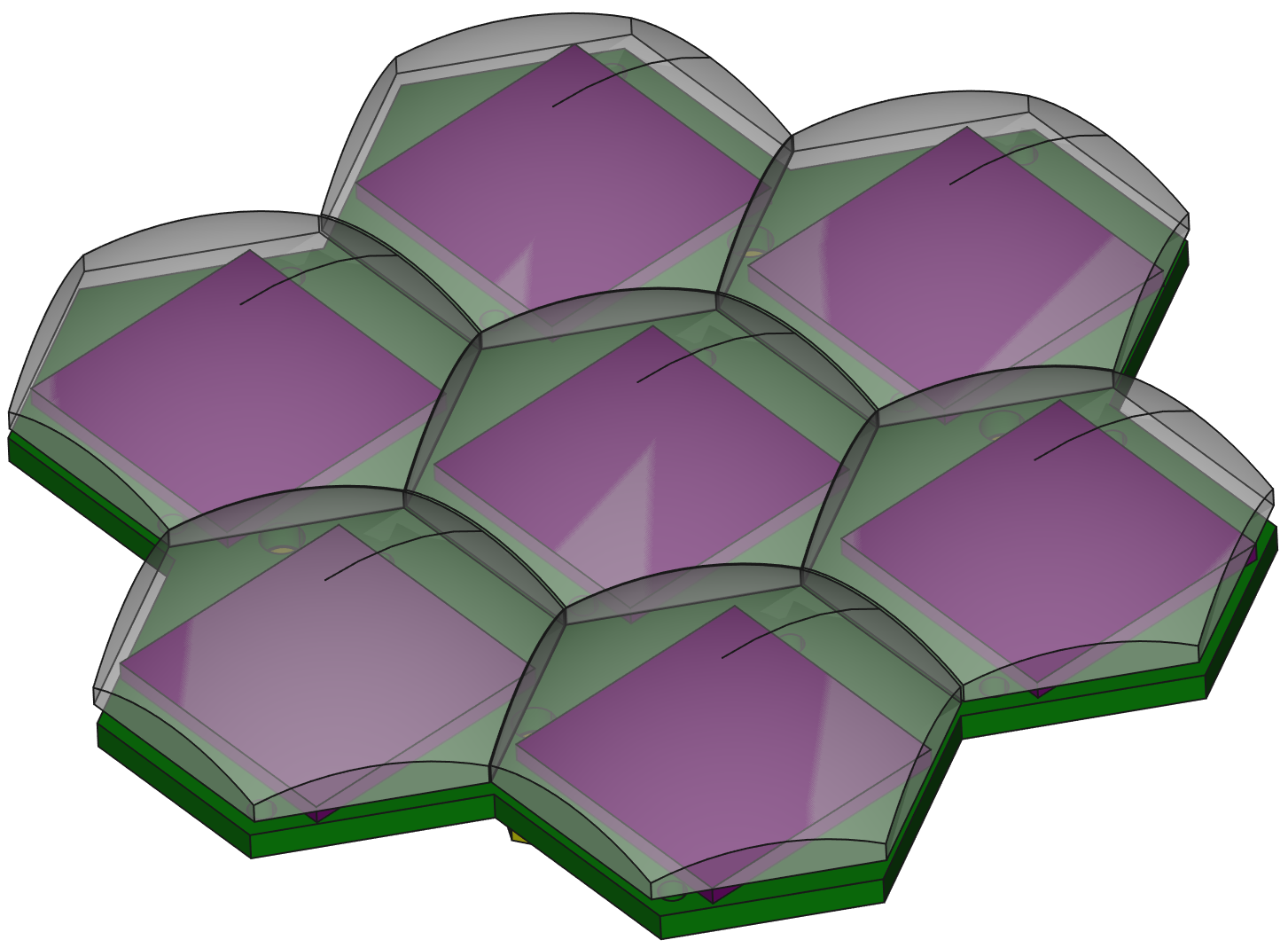} 
\captionstyle{normal} \caption{(Left) SPHERE-3 optical design. (Right) General view SiPM segment with lens light collectors. Purple squares represent $6\times6$ mm SiPMs.}
\label{fig:optics}
\end{figure}

A diagram of the optical setup is provided in Fig.~\ref{fig:optics} left. All structural components of the system are illustrated in Fig.~\ref{fig:genview}, which shows them installed within a metal framework constructed from thin-wall aluminum tubes. Light sensitive camera (SiPM mosaic) is composed of 379 individual units, with seven silicon photomultipliers (SiPM) per unit (see Fig.~\ref{fig:optics} right). The separation between adjacent SiPM centers is approximately 12~mm. The mosaic is positioned at the focus of the mirror, at a distance of 592~mm. 

The measuring equipment is installed directly beneath the photosensitive mosaic. This facilitates the installation process and minimizes the length of cables, thereby reducing the overall mass of the detector.

To prevent overheating of the SiPM array due to heat dissipation from the measurement system boards, cooling fans are mounted in the lower section of the electronics module. Cold air enters through slots between the SiPM units, passes through the electronics, and is expelled from below, beneath the corrective lens.

Power for the electronics module is supplied from the UAV power system via internal cables within the detector frame. In the upper part of mirror a direct light detector is planned (however, this placement and direct CL detector design is not finalized). 
Few options for this detector design were under consideration:

A) An array of pinholes forming a coding aperture in the upper portion of the mirror. During the research, it was determined that, with the size of these apertures corresponding to the pixel size (or module size), there are too few photons from direct CL entering a single pixel to successfully reconstruct angular distributions. As the number and size of the pinholes increase, the quality of recording of reflected CL begins to deteriorate. This design may function in the absence of background photons scattered from snow and collected by the mirror, but will require a shutter of sorts on the entrance window. Therefore, at this point, it was decided to discontinue this version.

B) A lens installed in the mirror center to form an image on the camera surface. The mirror's center is fully shaded by the camera and does not participate in reflected light collection. But the resolution of this detector proved to be too low for any useful data to be obtained from direct CL distribution. Since the resolution was limited primarily by the low pixel count in detector camera this option was also dropped.

C) Instead, a separate detector with its own optical system and a mosaic based on either a CCD matrix or a matrix of individual SiPMs with dimensions of 1x1 or 3x3 mm was proposed. The field of view for the telescope would be a cone with half-angle solutions of 15--20$^\circ$ for a single-lens design. The functionality of the detector assembly is currently under investigation. It comprises a collecting lens with a surface area of 400 cm$^2$ and a focal length of 63.5 cm, and a sensor positioned at the focal point.

D) Another option is a set of 7 lenses 400 cm$^2$ each arranged into hexagonal pattern with $\sim1.2$~m focal length and a sensitive camera built from 16 (4 by 4 square) 8-by-8 SiPM modules. Optical resolution of this design is around 0.19--0.21$^\circ$. This option has same field of view as the previous one.

Two last options are now in careful study. Option C does not require much of the modelling to be done, but requires the search for suitable UV sensitive CCD matrix with extremely large pixel or a matrix of small UV-sensitive SiPMs. None of these components are readily available for purchase at the moment. Option D involves significantly more steps to be added to the simulations pipeline~\cite{Ivanov2024}.

Tab.~\ref{tab:prop} presents a summary of the SPHERE-3 system and its prototype parameters, as well as the characteristics of the previous SPHERE-2 system for comparison. The prototype serves as an intermediate step in the development process of the project, and will be used for conducting experimental tests on individual components of the measurement system, such as SiPM segments, FADC boards, and trigger etc. Additionally, algorithms for interacting with unmanned aerial vehicles (UAVs) will be developed during this phase.

\begin{table}[!ht]
\setcaptionmargin{0mm}
\onelinecaptionstrue
\captionstyle{flushleft}
\caption{SPHERE-2 and SPHERE-3 parameters}
\label{tab:prop}
\bigskip
\begin{tabular}{|l|c|c|c|}
  \hline
   Parameter & SPHERE-2 & Prototype & SPHERE-3 \\
  \hline
  Effective entry window area, m$^2$ & 0.50 & 0.16 & 1.92 \\
  Mirror diameter, m & 1.5 & 0.8 & 2.2 \\
  Viewing angle & $\pm25^\circ$ & $\pm20^\circ$ & $\pm20^\circ$ \\
  Number of pixels & 109 & 259 & 2653 \\
  Detector weight, kg & 90 & 15 & 100 \\
  Maximum altitude, m & 900 & 500 & 1500 \\
  Carrier & balloon & UAV & Heavy UAV \\  [1mm]
  \hline
\end{tabular}
\end{table}

\section{Events registration}
\subsection{Reflected CL trigger}
To detect the EAS CL in the SPHERE-3 telescope, a two-step triggering algorithm is implemented~\cite{Entina2024}. The first step is an improved topological trigger, which is activated when the adaptive threshold in a group of neighboring pixels is exceeded. This algorithm is similar to the one used in the previous version of the SPHERE telescope (SPHERE-2)~\cite{Antonov2019}. 

However, when using SiPMs, there is an increased likelihood of multiple random triggers due to cross-talks producing high amplitudes in response to a single photon, leading to a significant increase in the number of false triggers. To mitigate this issue, a second stage of processing is employed --- an algorithm based on a convolutional neural network that analyzes fragments of the detector buffer data to detect characteristic spatial patterns.

To train the neural network, synthetic data sets were generated, including simulations of air showers using the CORSIKA code, accounting for atmospheric and optical effects, as well as the response of the electronic circuitry of the detector. The use of such a system allows for a reduction in the frequency of false positives by several orders of magnitude, and consequently, a lowering of the registration threshold.

\subsection{Direct CL trigger}

To implement the direct light trigger algorithm, the Geant4-based direct light detector simulation program was modified to take into account the temporal structure of photon interaction with the detector. After the simulation, the coordinates of the detected photons, initially determined in the system associated with the shower axis, are transformed into the terrestrial coordinate system. Additionally, background photons are estimated, modelled on the basis of empirical parameters characterizing the background illumination: 
\[I \cdot S \cdot \Omega \cdot t \approx 182 \, \text{ photons},\]
where 
$ I = 4 \cdot 10^{12} \, \text{ photon}\cdot\text{m}^{-2} \cdot \text{s}^{-1} \cdot \text{sr}^{-1}$  --- averaged ground illumination level from night sky;
$ S = 400 \, \text{ cm}^2 = 0.04 \, \text{ m}^2 $ --- detector effective area;
$\Omega(20^\circ) = 2\pi (1 - \cos(\phi/2)) = 0.095 \, \text{ sr} $ --- detector field of view sold angle;
$ t \sim 12.5 \, \text{ ns} $ --- preliminary time bin length (i.e. inverse smapling frequency).

Events and background triggers are separated by exceeding the threshold value by at least one pixel of the frame; the threshold value is determined statistically by the intensity distribution in the training sample.

\begin{figure}[t!]
\setcaptionmargin{5mm}
\onelinecaptionstrue  
\includegraphics[width=0.49\linewidth]{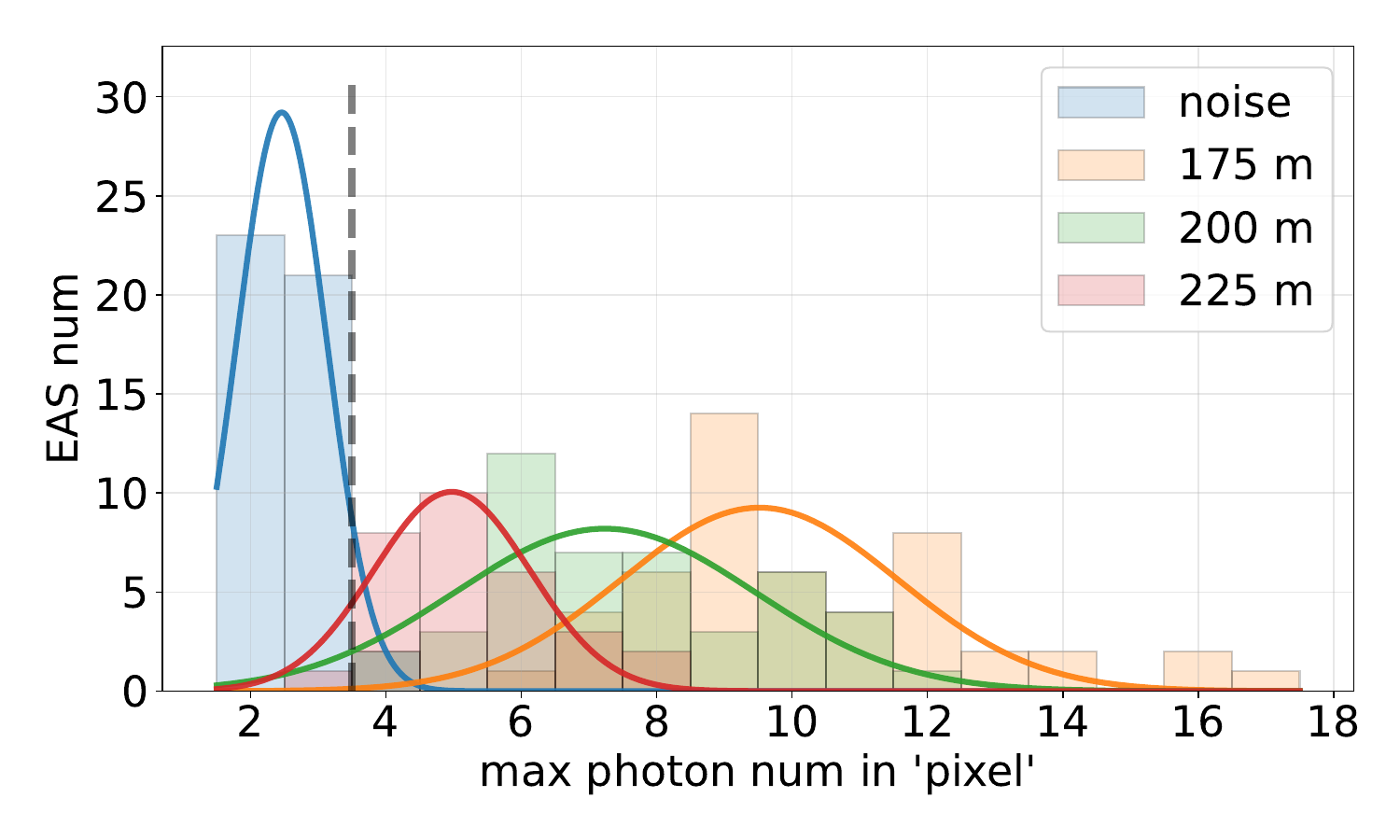}
\captionstyle{normal} \caption{Distributions of the maximum number of photons per pixel for background photons (blue) and for EAS events (proton, 1 PeV, 5$^\circ$ zenith angle) with trajectories at a distance of 175 (orange), 200 (green) and 225 (red) m from the telescope. The gray vertical line illustrates the filtering threshold.}
\label{fig:pix_trig}
\end{figure}

Fig.~\ref{fig:pix_trig} shows the distributions of photon counts in pixels with the maximum signal for the 1~PeV EAS events, whose axis passed at a considerable distance from the telescope (175, 200, and 225 meters), and a similar distribution only for background photons. Since the amplitude of the signal significantly exceeds the background noise level, filtering by the maximum pixel value ensures high accuracy of event selection.

\section{Primary particle type estimation}

At this stage, a generalized version of the angular image detector is being used. The numerical model for the direct light detector was developed using Geant4 and comprises a lens and a sensor array with a 100\% efficiency. The spatial and angular distributions of Cherenkov light at the flight altitude were modelled using the CORSIKA code, and these distributions served as the light source for the simulation. The passage of optical photons through the detector was simulated using Geant4 toolkit, and the output is a list of photons indicating the coordinates of their absorption point by the sensor.

For a complex mass assessment, it is essential that the event be within the field of view of both direct and indirect light detectors. Each such event corresponds to a point in a two-dimensional parameter space: the major axis length for direct CL images and the ratio of integral values along the inner and outer circles for reflected CL. Similar to the simple direct CL mass estimation method, this approach takes into account the distance between the detector and the shower axis as well as the relative azimuthal position of the detector. A dividing line is then selected to minimize classification errors. Currently, the use of complex mass assessments allows us to achieve classification errors as listed in the final row of Table~\ref{tab:qual}. The `indirect CL' row contains classification errors from indirect light mass estimations, and the `direct CL' row indicates errors from direct light measurements. As can be seen, the complex approach allows for the best possible results. A more accurate method for determining the mass of a particle using direct light is currently under development, and we anticipate that incorporating this new criterion into a complex approach will further minimize classification errors.

\begin{table}[!ht]
\setcaptionmargin{0mm}
\onelinecaptionstrue
\captionstyle{flushleft}
\caption{Preliminary results of the mass estimation using dual classification.}
\label{tab:qual}
\bigskip
\begin{tabular}{|l|c|c|c|c|}
  \hline
   \multirow{2}{*}{} & \multicolumn{2}{c|}{p-N separation} & \multicolumn{2}{c|}{N-Fe separation}\\
   \cline{2-5}
   & error p & error N & error N & error Fe \\  
   \hline
   reflected CL      & 0.32 & 0.32 & 0.31 & 0.31 \\
   direct CL         & 0.25 & 0.26 & 0.23 & 0.24 \\
   combined analysis & 0.22 & 0.15 & 0.19 & 0.14 \\ [1mm]  
  \hline
\end{tabular}
\end{table}

\section{Conclusion}
Work is underway to improve the method for the detection of the reflected Cherenkov light
A preliminary design of the detector and optical system for the SPHERE-3 detector and its prototype has been developed. High sensitivity of the direct Cherenkov light to the primary mass was demonstrated. Method of the primary mass estimation using the data of both detectors was created

\begin{acknowledgments}
This work is supported by the Russian Science Foundation under Grant No. 23-72-00006,
https://rscf.ru/project/23-72-00006/. The research was carried out using the equipment of the shared research facilities of HPC computing resources at Lomonosov Moscow State University~\cite{SC}.
\end{acknowledgments}

%

%


\end{document}